\documentclass[intlimits,twoside,a4paper]{article}
\usepackage[cp1251]{inputenc}
\usepackage[eqsecnum]{cmpj3}
\usepackage{bm}

\issue{2021}{24}{2}{23701}
\doinumber{10.5488/CMP.24.23701}

\title[Current-phase relation in layered superconducting structures of SIS’IS type]
{Current-phase relation in layered superconducting structures of SIS’IS type}

\author[A. M. Shutovskyi, V. E. Sakhnyuk]{A. M. Shutovskyi\orcid{0000-0001-7628-0078}, V. E. Sakhnyuk\orcid{0000-0002-9225-7473}}
\address{Lesya Ukrainka Volyn National University, 13 Voli Ave., 43000 Lutsk, Ukraine}

\Keywords{Josephson effect, Josephson junctions, Green's function methods}

\date{Received November  26, 2020, in final form February  1, 2021}
\begin{document}

\maketitle

\begin{abstract}
The dependence of the current density on the phase difference is investigated considering the layered superconducting structures of a SIS’IS type. To simplify the calculations, the quasiclassical equations for the Green’s functions in a \textit{t}-representation are derived. An order parameter is considered as a piecewise constant function. To consider the general case, no restrictions on the dielectric layer transparency and the thickness of the intermediate layer are imposed. It was found that a new analytical expression for the current-phase relation can be used with the aim to obtain a number of previously known results arising in particular cases.
\printkeywords
\end{abstract}

\section{Introduction}

In the microscopic theory of superconductivity, a layered superconducting structure of a SIS’IS type (S and S’ are the superconductors, I is an insulator) is also called a double-barrier superconducting junction. For certain values of the intermediate layer thickness, the so-called resonant tunneling modes~\cite{r09,n99} can arise in the layered superconducting structures of a SIS’IS type. In paper~\cite{b00}, an interesting feature of the supercurrent flow is investigated considering the layered superconducting structures of a SIS’IS type. Using a microscopic approach, the features of the charge transport in a three-dimensional SIS’IS junction are theoretically investigated in the absence of nonmagnetic impurities. Depending on the thickness of the intermediate layer, there are two electron transport modes in the layered superconducting structures of a SIS’IS type. The first mode is also called a coherent mode. In the case of a coherent mode, the supercurrent of the junction is directly proportional to the dielectric layer transparency $D$. When the thickness of the intermediate layer is increased, there is a mode with a broken coherence. This is the second mode. In the case of the second mode, the supercurrent of the junction becomes directly proportional to $D^2$.

Considering a double-barrier Josephson junction as a coupled three-level quantum system, a Feynman-like method~\cite{f65,b82} can be generalized for the layered superconducting structures of a SISIS type. It is significant to note that the three superconductors have the same Cooper pair density~\cite{m02}. In the paper~\cite{c96}, a modified Ohta’s model~\cite{o76} is applied with the aim to explore the stationary properties of the SISIS junctions.

The effects of the phase coherence in the layered superconducting structures are also important in terms of practical application~\cite{s06,p00,l13,n94,k99}. This is experimentally demonstrated by the possibility to construct the Josephson junctions with desirable engineering properties using the known layering methods~\cite{n999,s98}.

A double-barrier Josephson junction of a SIS’IS type is fabricated from Nb/Al-AlO$_{x}$-Nb/Al-AlO$_{x}$-Nb/Al structures deposited in the same vacuum with the aim to provide the identical material parameters of the films and the tunnel barriers. In the paper~\cite{n97}, the fabrication procedure is described more in detail. In the paper~\cite{n99}, an absolute value of a Josephson current in a double-barrier Josephson junction with very thin middle electrodes is studied experimentally. The two junctions are separated by the distance of an order of a superconducting coherence length. It is also concluded that the maximum value $J_c$ of the current density in a Josephson junction of a SIS’IS type is higher than the corresponding maximum value $J_c$  in a single S’IS junction from the stack. This effect is explained by a model that takes an Andreev reflection~\cite{a66} in a SIS’IS double-barrier stack into consideration.

The main purpose of our paper is to explore the stationary properties in the layered superconducting structures of a SIS’IS type. An order parameter of the superconductor S’ is different from an order parameter of the superconductor S. Unlike the above-mentioned papers, a general case will be considered in our investigation. This means that no restrictions on the dielectric layer transparency and the thickness of an intermediate layer are imposed. The main task of the investigation is to obtain the dependence of the supercurrent on the phase difference. The dependence of the supercurrent on the thickness of the intermediate layer S’ must also be analyzed.

A research is conducted using a Green’s function method~\cite{s82}. To obtain an analytical expression for the Green’s function, a spatial smoothing procedure on the length scales of an order of an atomic size is performed. This enables us to derive the quasiclassical equations describing the spatial behavior of the Green’s functions on the length scales of an order of the superconducting coherence length. It is shown that the most compact form of quasiclassical equations can be found using a $t$-representation~\cite{s82}. In this representation, an expression for the current density can also be obtained. Considering an order parameter as a piecewise constant function, the solutions of the quasiclassical equations for the Green’s functions are found in the three superconductors. The absolute value of an order parameter is equal to $\Delta_1$ for a superconductor of the intermediate layer. The absolute value of an order parameter is equal to $\Delta$ for other superconductors.

The proposed calculation scheme enables us to obtain a new analytical expression for the dependence of the current density on the phase difference. A current-phase relation is true for arbitrary values of the dielectric layer transparency and an intermediate layer thickness. A new analytical expression for the current-phase relation also enables us to obtain a number of previously known results for the following junctions: SNS, SINIS, SISIS and SIS. Thus, a thorough verification of the correctness result is performed.

\section{Model and basic equations}

Let a plane XOY be a middle of an intermediate layer S’. The thickness of an intermediate layer is equal to $d$. The insulator films are placed on the planes $z=\pm \left( {d}/{2} \right)$. The left-hand superconductor and the right-hand superconductor have a location $|z|>{d}/{2}$. On the plane XOY, there is a translational invariance. The spatial homogeneity is broken in the direction of the OZ axis. It is significant to note that the insulator films can be modelled by a potential field
\[
U(z)=\frac{U_0}{2}\left[\delta\left(z-\frac{d}{2}\right)+\delta\left(z+\frac{d}{2}\right)\right], \quad U_0>0.
\]

In the microscopic theory of superconductivity, the Matsubara Green’s functions are considered to be a powerful mathematical tool. The thermodynamic quantities of a superconductor can be represented via the Matsubara Green’s functions. It is significant to note that these Green’s functions~\cite{s82} are the elements of a matrix
\begin{align}
\hat{G}_{\omega_n}\left(\vec{r},\vec{r}'\right)=\left(
\begin{array}{cc}
G_{\omega_n}\left(\vec{r},\vec{r}'\right) & -\tilde{F}_{\omega_n}\left(\vec{r},\vec{r}'\right)\\
-F_{\omega_n}\left(\vec{r},\vec{r}'\right) & \tilde{G}_{\omega_n}\left(\vec{r},\vec{r}'\right)\\
\end{array}
\right).
\label{m1}
\end{align}
Let us consider the absence of an external magnetic field $\vec{A}(\vec{r})$ and the presence of a potential field $U(z)$. In this particular case, a matrix Green’s function~(\ref{m1}) is a general solution of the second order differential equation
\begin{align}
\left[ \ri\omega_n-\sigma_z\left(\hat{\xi}+U(z)\right)-\hat{\Delta}(z)\right]
\hat{G}_{\omega_n}\left(\vec{r},\vec{r}'\right)=
\delta(\vec{r}-\vec{r}').
\label{m2}
\end{align}

In the differential equation~(\ref{m2}), there are the odd Matsubara frequencies $\omega_n=\piup T (2n+1)$  dependent on the numbers $n\in Z$. We also have a Pauli matrix $\sigma_z=\left(\begin{smallmatrix} 1&0\\ 0 & -1\\ \end{smallmatrix} \right)$,  a differential operator
$\hat{\xi}=\frac{\hat{\vec{p}}^2}{2m}-\mu$
and a matrix   $\hat{\Delta}(z)=\left(\begin{smallmatrix} 0 & \Delta(z)\\ \Delta^*(z) & 0\\ \end{smallmatrix} \right)$
containing an order parameter $\Delta(z)$.

In the superconductor, the motion of the Cooper pairs can be considered to be a quasiclassical motion, because the Fermi momentum $p_0$ is much greater than the momentum $mv_s$ related to the motion of a Cooper pair. Such a conclusion can be made considering an estimator of a physical quantity $v_s^c \sim {\Delta}/{p_0}\sim {T_c}/{p_0}$. As a result, there is a fraction ${m v_s}/{p_0}\ll {T_c}/{E_{\textrm{F}}}\sim {a}/{\xi_0}$ containing an atomic length $a$. In the theory of low temperature superconductivity, the above-mentioned fraction is an infinitesimal parameter. As a result, a spatial smoothing procedure on the length scales of an order of an atomic size can be performed for an order parameter and some other functions. This enables us to neglect the small-scale changes. Only the large-scale changes on the length scales of an order of a superconducting coherence length must be taken into account. The second order differential equations for the Green’s functions can be transformed into the first order differential equations by a spatial smoothing procedure. We are now talking about the so-called quasiclassical equations. In the book~\cite{s82}, a way to derive the quasiclassical equations is demonstrated for the superconducting junctions of an SIS type. In our case, the way to derive the quasiclassical equations for the layered superconducting structures of an SIS’IS type is analogous to the way demonstrated in a book~\cite{s82}. Considering the layered superconducting structures of an SIS’IS type, it was found that the first order differential equations for the Green’s functions in a \textit{t}-representation are analogous to the first order differential equations for the superconducting junctions of an SIS type.

\section{Quasiclassical equations}

Let us consider the second order differential equation
\begin{align}
\left[\hat{\xi}+U(z)\right]\psi_{\vec{p}}^{(k)}(\vec{r})=\xi_{\vec{p}} \psi_{\vec{p}}^{(k)}(\vec{r})
\label{m3}
\end{align}
containing a dispersion relation $\xi_{\vec{p}}=\frac{p^2}{2m}-\mu$. A superscript $k$  can acquire values of 1 and 2. A general solution of the second order differential equation~(\ref{m3}) is a three-dimensional wave function
\begin{align}
\psi_{\vec{p}}^{(k)}(\vec{r})=\frac{1}{2 \piup}\exp(\ri \vec{p}_{\perp}\vec{r})\chi_{p_z}^{(k)}(z).
\label{m4}
\end{align}

In the formula~(\ref{m4}), a notation for the so-called transverse momentum $\vec{p}_{\perp}=p_x\vec{e}_x+p_y\vec{e}_y$    is introduced. Substituting the three-dimensional wave function~(\ref{m4}) into a second order differential equation~(\ref{m3}), we can derive a one-dimensional second order differential equation with the following solutions:

\begin{align}
\chi_{p_z}^{(1)}(z)&=\frac{1}{\sqrt{2\piup}}\bigg[
\frac{C_1}{2}\re^{-\ri p_zz}+\ri (C_2-1)\re^{-\ri p_z\frac{d}{2}}\textrm{sign}\left(z+\frac{d}{2}\right)
\sin p_z\left(z+\frac{d}{2}\right)\nonumber\\
&+\ri C_3\re^{-\ri p_z\frac{d}{2}}\textrm{sign}\left(z-\frac{d}{2}\right)\sin p_z\left(z-\frac{d}{2}\right)+
\frac{1+C_4}{2}\re^{\ri p_z z}
\bigg], \quad \chi_{p_z}^{(2)}(z)=\chi_{p_z}^{(1)}(-z).
\label{m5}
\end{align}

In the formula~(\ref{m5}), there is a momentum $p_z>0$. We also have constant coefficients defined by the formula
$C_1=(C_2-1)\re^{-\ri p_z d}+C_3$, $C_2=\left(1+\frac{\ri K}{2 p_z}\right)C_4$, $C_3=-\frac{\ri K}{2 p_z}C_4 \re^{\ri p_z d}$ and
\begin{align}
C_4=\frac{\re^{-\ri p_z d}}{\left(1+\frac{\ri K}{2 p_z}\right)^2 \re^{-\ri p_z d}+\left(\frac{K}{2 p_z}\right)^2 \re^{\ri p_z d}}.
\nonumber
\end{align}

Here, we have a notation $K=m U_0$. The coefficient $C_4$  can be used with the aim to calculate the dielectric layer transparency
\begin{align}
D=|C_4|^2=\left[1+\left(\frac{K}{p_z}\right)^2\left(\cos p_z d+\frac{K}{2p_z}\sin p_z d\right)^2\right]^{-1}
\nonumber
\end{align}
and the reflection coefficient $R=|C_1|^2=1-D$. A three-dimensional wave function~(\ref{m4}) can be used with the aim to obtain an expansion of the matrix Green’s function~(\ref{m1}). Thus,
\begin{align}
\hat{G}_{\omega_n}(\vec{r},\vec{r}')=
\sum_{i,k}\int\rd \vec{p} \int\rd \vec{p}' \hat{G}_{\omega_n}^{i,k}(\vec{p},\vec{p}')
\psi_{\vec{p}}^{(i)}(\vec{r}) \psi_{\vec{p}'}^{*(k)}(\vec{r}').
\label{m6}
\end{align}

The expansion~(\ref{m6}) must be substituted into the second order differential equation~(\ref{m2}). This enables us to derive the equations for the Green's functions in the momentum representation 
\begin{align}
 \left( \ri{{\omega }_{n}}\!-\!{{\sigma }_{z}}\xi  \right)
 \hat{G}_{{{\omega }_{n}}}^{i,k}\left( \vec{p},{\vec{p}}' \right)-\!\sum\limits_{{{i}'}}\!{\int\!\!{\rd{\vec{p}}''\left[ \int\!\!{\rd\vec{r}{{\psi }^{*\left( i \right)}_{{\vec{p}}}}\left( {\vec{r}} \right)\hat{\Delta }\left( z \right)\psi _{{{\vec{p}}''}}^{\left( {{i}'} \right)}\left( {\vec{r}} \right)} \right]\!\hat{G}_{{{\omega }_{n}}}^{{i}',k}\left( {\vec{p}}''\!\!,{\vec{p}}' \right)}}={{\delta }_{i,k}}\delta \left( \vec{p}\!-\!{\vec{p}}' \right).
\label{m6_1}
\end{align}

In fact, the spatial homogeneity is broken only in the direction of the axis OZ. Thus, the order parameter $\Delta$ depends only on the coordinate $z$. As a result, the matrix elements $\hat \Delta$ contain the function $\delta(\vec p_{\bot}-{\vec p}''_{\bot})$. The Green's function in the momentum representation has the same property 
$ \hat G^{i,k}_{\omega_n}(\vec p,\vec p')=\hat G^{i,k}_{\omega_n}(\vec p_{\bot},p_z,p_z')
\delta(\vec p_{\bot}-\vec p'_{\bot}). $
According to the above-mentioned, the equation for the Green's function in the momentum representation can be presented as follows:
\begin{align}
  & \left( \ri{{\omega }_{n}}-{{\sigma }_{z}}\xi  \right)\hat{G}_{{{\omega }_{n}}}^{i,k}\left( {{{\vec{p}}}_{\bot }},{{p}_{z}},{{{{p}'_{z}}}} \right)\nonumber\\
   &-\sum\limits_{{{i}'}}{\int\limits_{0}^{+\infty }{\rd{{{{p}''_{z}}}}\left[ \int\limits_{-\infty }^{+\infty }{\rd z{{\chi }^{*\left( i \right)}_{{{p}_{z}}}}\left( z \right)\hat{\Delta }\left( z \right)\chi _{{{{{p}''_{z}}}}}^{\left( {{i}'} \right)}\left( z \right)} \right]\hat{G}_{{{\omega }_{n}}}^{{i}',k}\left( {{{\vec{p}}}_{\bot }},{{{{p}''_{z}}}},{{{{p}'_{z}}}} \right)}}={{\delta }_{i,k}}\delta \left( {{p}_{z}}-{{{{p}'_{z}}}} \right).
  \label{m6_3} 
\end{align}
It is well known that the characteristic values of the momentum are close to the Fermi momentum $p_{0}$. Hence, the identity
$\displaystyle\xi={p^{2}}/{2m}-{p_{0}^{2}}/{2m}\cong {{v}_{0}}\left( p-{{p}_{0}} \right)$
can be used with the aim to derive the following identities: 
$\displaystyle p=p_{0}+{\xi}/{v_{0}}$, $\displaystyle p'=p_{0}+{\xi'}/{v_{0}}.$ 
Using the conservation law for the transverse momentum $\vec{p}'_{\bot}=\vec{p}_{\bot}$ or $p\sin\theta =p'\sin{\theta}'$, we obtain the following relationships between the angles: 
$$
\begin{matrix}\displaystyle
\sin\theta'\cong\left( 1+\frac{\xi-\xi'}{p_{0}v_{0}}\right)\sin\theta,
&
\displaystyle
\cos\theta'\cong\cos\theta-\frac{\xi-\xi'}{p_{0}v_{0}}\sin\theta\tan\theta.
\end{matrix}
$$
The derived identities are linear functions dependent on the difference $\xi-\xi'$. The approximation for the difference
$p_{z}-p'_{z}$ can be derived from the relation
$$
\xi-\xi'=\frac{1}{2m}\left(p_{z}-p'_{z}\right)\left(p_{z}+p'_{z}\right)\cong{v_{0}}\cos\theta\left(p_{z}-p'_{z}\right).
$$
Thus, there are the approximations
$\displaystyle p'_{z}-p_{z}\cong\frac{\xi'-\xi}{v_{0}x}$
and
$\displaystyle \text{d}p'_{z}\cong\frac{\text{d}\xi'}{v_{0}x}.$
Here, we have the following notation:
$
\cos\theta\equiv x,\,\,
0<x<1.
$
The derived approximations are used with the aim to derive the equations for the Green's functions dependent on the variables $\xi$ and $\xi'$.
Then, the equation~(\ref{m6_3}) can be transformed into the equation
\begin{align}
&\left( \ri{{\omega }_{n}}-{{\sigma }_{z}}\xi  \right)\hat{G}_{{{\omega }_{n}}}^{i,k}\left( \xi ,{\xi }' \right)\nonumber\\
&-\frac{1}{{{v}_{0}}x}\int\limits_{-\infty }^{+\infty }{\rd{\xi }''\sum\limits_{{{i}'}}{\left[ \int\limits_{-\infty }^{+\infty }{\rd z{{\chi }^{*\left( i \right)}_{{{p}_{z}}}}\left( z \right)\hat{\Delta }\left( z \right)\chi _{{{{{p}''_{z}}}}}^{\left( {{i}'} \right)}\left( z \right)} \right]\hat{G}_{{{\omega }_{n}}}^{{i}',k}\left( {\xi }'',{\xi }' \right)}}={{v}_{0}}x{{\delta }_{i,k}}\delta \left( \xi -{\xi }' \right).
\label{m6_4} 
\end{align}

Then, we perform an inverse Fourier transformation into the space of variables conjugate to $ \xi $ and $ {\xi} '$. To do this, we introduce a full set of functions:
$\frac{{{\re}^{\ri\xi t}}}{{{(2\piup {{v}_{0}}x)}^{1/2}}},\ \ \frac{{{\re}^{-\ri\xi t}}}{{{(2\piup {{v}_{0}}x)}^{1/2}}}$.
The equation~(\ref{m6_4}) should be multiplied by 
$
\frac{1}{2\piup v_{0}x}\exp \left({\ri\xi t-\ri\xi't'}\right).
$
The obtained equation should be integrated over the variables $\xi$ and $\xi'$.
As a result, we can derive the equations for the Green's functions in the $t$-representation 
	\begin{align}
\left( \ri \omega_n+\ri \sigma_z \frac{\partial}{\partial t} \right)\hat{G}_{{{\omega }_{n}}}^{i,k}\left( t,{t}' \right)-\int\limits_{-\infty }^{+\infty }{\rd{t}''\sum\limits_{{{i}'}}{\left\langle  t,i \right|\hat{\Delta }\left( z \right)\left| {i}',{t}'' \right\rangle \hat{G}_{{{\omega }_{n}}}^{{i}',k}\left( {t}'',{t}' \right)}}={{\delta }_{i,k}}\delta \left( t-{t}' \right).
	\label{m6_5} 
\end{align}
Here, we have the notation
	\begin{align*}
	 \hat{G}_{{{\omega }_{n}}}^{i,k}\left( t,{t}' \right)&=\frac{1}{2\piup {{v}_{0}}x}\int\limits_{-\infty }^{+\infty }{\rd\xi \int\limits_{-\infty }^{+\infty }{\rd{\xi }'\hat{G}_{{{\omega }_{n}}}^{i,k}\left( \xi ,{\xi }' \right){{\re}^{\ri\xi t-\ri{\xi }'{t}'}}}},	\\
	\left\langle t,i\right|\hat{\Delta}\left(z\right)\left|{i}',{t}'\right\rangle &=\frac{1}{2\piup{v_{0}}x}\int\limits_{-\infty }^{+\infty }{\rd\xi \int\limits_{-\infty }^{+\infty }{\rd{\xi }'\left( \int\limits_{-\infty }^{+\infty }{\rd z\hat{\Delta }\left( z \right){{\chi }^{*}}_{{{p}_{z}}}^{\left( i \right)}\left(z\right)\chi_{p'_{z}}^{\left(i'\right)}\left(z\right)}\right){{\re}^{\ri\xi t-\ri\xi't'}}}}.
	 \end{align*}	
	 Calculating the matrix elements $\left\langle  t,i \right|\hat{\Delta }\left( z \right)\left| {i}',{t}'' \right\rangle $, we obtain the following property:
	$\left\langle  t,i \right|\hat{\Delta }\left( z \right)\left| {i}',{t}'' \right\rangle =\delta \left( {t}''-t \right){{\hat{\Delta }}^{i,{i}'}}\left( t,x \right).$
This enables us to evaluate the integral over the variable ${t}''$ in the equation~(\ref{m6_5}). As a result, we can derive the quasiclassical equations for the Green's functions in the $t$-representation
\begin{align}
\left( \ri \omega_n+\ri \sigma_z \frac{\partial}{\partial t} \right)
\hat{G}_{\omega_n}^{i,k}(t,t')-\sum_{i'}\hat{\Delta}^{i,i'}(t,x)
\hat{G}_{\omega_n}^{i',k}(t,t')=\delta_{i,k}\delta(t-t').
\label{m7}
\end{align}

To calculate the matrix functions ${{\hat{\Delta }}^{i,{i}'}}\left( t,x \right)$, the spatial smoothing procedure on the length scales of an order of an atomic size must be performed. Considering the exponents ${\exp \left[{\ri\left( {{p}_{z}}+{{{{p}'}}_{z}} \right)z}\right]}$ and ${\exp \left[{\ri\left( {{p}_{z}}-{{{{p}'}}_{z}} \right)z}\right]}$, we may note that the exponents of the first type are approximately equal to ${\exp \left({2\ri{{p}_{0}}\cos \theta z}\right)}$ and oscillate on the length scales of an order of an atomic size. In the exponents of the second type, the large values of the Fermi momentum are annihilated. As a result, these exponents are approximately equal to ${\exp \left\lbrace {\ri\left[\left({\xi -{\xi }'}\right)/{{{v}_{0}}x}\right]z}\right\rbrace}$ and oscillate on the length scales of an order of the superconducting coherence length ${{\xi }_{0}}\gg a$. The goal of our investigation is to take only the large-scale changes into consideration. Thus, the exponents of the first type should be neglected. 

As a result, we obtain the following formulae:
\begin{align}
\hat{\Delta}^{1,1}(t,x)&=\theta\left(-\frac{a}{2}-t\right)\hat{\Delta}(v_0xt)+
\left[\theta\left(t+\frac{a}{2}\right)-\theta\left(t-\frac{a}{2}\right)\right]
\left[|C_2|^2 \hat{\Delta}(v_0xt)+|C_3|^2 \hat{\Delta}(-v_0xt) \right]\nonumber\\
&+\theta\left(t-\frac{a}{2}\right)\left[R \hat{\Delta}(-v_0xt)+D \hat{\Delta}(v_0xt)  \right],\nonumber\\
\hat{\Delta}^{1,2}(t,x)&=\left[\theta\left(t+\frac{a}{2}\right)-\theta\left(t-\frac{a}{2}\right)\right]
\left[C_2^*C_3\hat{\Delta}(v_0xt)+C_3^*C_2\hat{\Delta}(-v_0xt)\right]\nonumber\\
&+C_1^*C_4 \theta\left(t-\frac{a}{2}\right)\left[\hat{\Delta}(-v_0xt)-\hat{\Delta}(v_0xt)\right],\nonumber\\
\hat{\Delta}^{2,1}(t,x)&=\left[\theta\left(t+\frac{a}{2}\right)-\theta\left(t-\frac{a}{2}\right)\right]
\left[C_3^*C_2\hat{\Delta}(v_0xt)+C_2^*C_3\hat{\Delta}(-v_0xt)\right]\nonumber\\
&+C_1^*C_4 \theta\left(t-\frac{a}{2}\right)\left[\hat{\Delta}(v_0xt)-\hat{\Delta}(-v_0xt)\right],\nonumber\\
\hat{\Delta}^{2,2}(t,x)&=\theta\left(-\frac{a}{2}-t\right)\hat{\Delta}(-v_0xt)+
\left[\theta\left(t+\frac{a}{2}\right)-\theta\left(t-\frac{a}{2}\right)\right]
\left[|C_2|^2 \hat{\Delta}(-v_0xt)+|C_3|^2 \hat{\Delta}(v_0xt) \right]\nonumber\\
&+\theta\left(t-\frac{a}{2}\right)\left[D \hat{\Delta}(-v_0xt)+R \hat{\Delta}(v_0xt)  \right].\nonumber
\end{align}

Here, we have a notation $a={d}/{v_0x}$ . A physical quantity $v_0={p_0}/{m}$  is called a Fermi velocity. The constant coefficients $C_1$, $C_2$,  $C_3$,  $C_4$, $R$ and $D$ are calculated using the approximations $p_z\cong p_0x$ and  $p_z'\cong p_0x$. In this paper, a spatial behavior of an order parameter is considered within the so-called model with a piecewise constant order parameter~\cite{s82}. This model is a widely applicable method to explore the superconducting junctions. In fact, this is the way to neglect the order parameter changes caused by the influence of the current density or the dielectric layer transparency. Considering the order parameter as a piecewise constant function, we may write down a formula
\begin{align}
\Delta(z)=\Delta\left[
\theta\left(-\frac{d}{2}-z\right)\re^{-\ri \frac{\varphi}{2}}+\theta\left(z-\frac{d}{2}\right)\re^{\ri \frac{\varphi}{2}}
\right]+
\Delta_1\left[
\theta\left(z+\frac{d}{2}\right)-\theta\left(z-\frac{d}{2}\right)
\right].
\label{m8}
\end{align}
However, for the temperature close to the critical, such a reduced model cannot
be used. In this region of temperature, the spatial nonhomogeneity of the order
parameter near IS interface must be taken into account~\cite{p11,p18}.

The approximation~(\ref{m8}) of the order parameter $\Delta(z)$ enables us to calculate the matrix
\begin{align}
\hat{\Delta}(z)&=\Delta \sigma_x\cos\frac{\varphi}{2}+\Delta \sigma_y\sin\frac{\varphi}{2}+\theta\left(z-\frac{d}{2}\right)
\left(\Delta \sigma_x\cos\frac{\varphi}{2}-\Delta \sigma_y\sin\frac{\varphi}{2}-\Delta_1\sigma_x \right)
\nonumber\\
&+\theta\left(z+\frac{d}{2}\right) \left(\Delta_1\sigma_x-\Delta \sigma_x\cos\frac{\varphi}{2}-\Delta \sigma_y\sin\frac{\varphi}{2} \right).
\label{m9}
\end{align}
Here, we have the Pauli matrices $\sigma_x=\left(\begin{smallmatrix}0&1\\1 & 0\\ \end{smallmatrix} \right)$ and $\sigma_y=\left(\begin{smallmatrix} 0&-\ri\\\ri & 0\\ \end{smallmatrix} \right)$. The matrix~(\ref{m9}) must be used with the aim to calculate the matrices $\hat{\Delta}(v_0xt)$,  $\hat{\Delta}(-v_0xt)$  and $\hat{\Delta}^{i,i'}(t,x)$. The equation (\ref{m2}) is the second order differential equation, whereas the equation~(\ref{m7}) is the first order differential equation. This is the main advantage of the so-called quasiclassical equations for the matrix Green’s functions in a $t$-representation. It is significant to note that the first order differential equation~(\ref{m7}) can be transformed into the first order differential equation
\begin{align}
\left( \ri \frac{\partial}{\partial t}+\ri \omega_n\sigma_z \right)
\hat{\mathcal{G}}^{i,k}(t,t')-\sum_{i'}\sigma_z\hat{\Delta}^{i,i'}(t,x)
\hat{\mathcal{G}}^{i',k}(t,t')=\delta_{i,k}\delta(t-t')
\label{m10}
\end{align}
for the matrix Green’s function  $\hat{\mathcal{G}}^{i,k}(t,t')=\hat{G}_{\omega_n}^{i,k}(t,t')\sigma_z$. In the configuration representation, there is a matrix

\begin{align}
\hat{\mathcal{G}}_{\omega_n}(\vec{r}, \vec{r}')=\left(\begin{array}{rr} G_{\omega_n}(\vec{r}, \vec{r}')&\tilde{F}_{\omega_n}(\vec{r}, \vec{r}')\\-F_{\omega_n}(\vec{r}, \vec{r}') & -\tilde{G}_{\omega_n}(\vec{r}, \vec{r}')\\ \end{array} \right).
\label{m11}
\end{align}
Considering the matrix~(\ref{m11}), we can see that the element in the first column of the first row is the Green’s function $G_{\omega_n}(\vec{r}, \vec{r}')$. It is significant to note that the current density $\vec{j}(\vec{r})$  can be represented via the Green’s function $G_{\omega_n}(\vec{r}, \vec{r}')$.

\section{Current density}

We have to find the current density expressed via the Green’s functions in a $t$-representation. The representation of the current density $\vec{j}(\vec{r})$   via a Green’s function $G_{\omega_n}(\vec{r}, \vec{r}')$  is a well-known formula 
\begin{align}
\vec{j}(\vec{r})=\frac{\ri e}{m}T\sum_{\omega_n}\lim_{\vec{r}'\rightarrow\vec{r}}(\nabla_{\vec{r}'}-\nabla_{\vec{r}})G_{\omega_n}(\vec{r}, \vec{r}').
\label{m12}
\end{align}
The expansion~(\ref{m6}) of the Green’s function $G_{\omega_n}(\vec{r}, \vec{r}')$   must be substituted into the expression for the current density~(\ref{m12}) with the aim to derive a formula $\vec{j}(\vec{r})=j(z)\vec{e}_z$ containing the notation
\begin{align}
j(z)=\frac{1}{2}\frac{e}{m}N(0)T\sum_{\omega_n}\int_0^1 \textrm{d}x
\int_{-\infty}^{+\infty}\textrm{d}t\int_{-\infty}^{+\infty}\textrm{d}t'
\sum_{i,k}G_{\omega_n}^{i,k}(t,t')I^{k,i}(x,t,t').
\label{m14}
\end{align}

Here, we have the electron density of states at the Fermi surface $N(0)={m^2v_0}/{2\piup^2}$  and the integrals
\begin{align}
I^{k,i}(x,t,t')=\int_{-\infty}^{+\infty}\textrm{d}\xi\int_{-\infty}^{+\infty}\textrm{d}\xi'\re^{-\ri \xi t+\ri \xi' t'}J^{k,i}_{p_z',p_z}(z)
\nonumber
\end{align}
containing the four quantities
\begin{align}
J^{k,i}_{p_z',p_z}(z)=\chi^{*(k)}_{p_z'}(z)\hat{p}_z \chi^{(i)}_{p_z}(z)-
\chi^{(i)}_{p_z}(z)\hat{p}_z \chi^{*(k)}_{p_z'}(z),
\nonumber
\end{align}
because the superscripts $i$  and $k$   can acquire values of 1   and 2. To calculate the quantities  $J^{k,i}_{p_z',p_z}(z)$, a spatial smoothing procedure on the length scales of an order of an atomic size must be performed. Only the large-scale changes should be taken into account. As a result, we consider only the exponents dependent on the difference  $p_z-p_z'\cong\left({\xi-\xi'}\right)/{v_0x}$. It is significant to note that we have to neglect the exponents dependent on the sum $p_z+p_z'$. The constant coefficients should be calculated using the approximations $p_z\cong p_0x$ and  $p_z'\cong p_0x$. The calculated quantities $J^{k,i}_{p_z',p_z}(z)$ should be used with the aim to calculate the integrals $I^{k,i}(x,t,t')$. After that, the calculated integrals $I^{k,i}(x,t,t')$  should be substituted into the expression for the one-dimensional current density~(\ref{m14}). Considering the one-dimensional current density $j(z)$ on the plane $z={d}/{2}$, we can derive a formula
\begin{align}
j\left(\frac{d}{2}\right)&=\piup e v_0 N(0)T\sum_{\omega_n}\int_0^1x \rd x\bigg[|C_2|^2G_{\omega_n}^{1,1}\left(\frac{a}{2},\frac{a}{2}\right)-
|C_3|^2G_{\omega_n}^{1,1}\left(-\frac{a}{2},-\frac{a}{2}\right)\nonumber\\
&+D G_{\omega_n}^{1,1}\left(\frac{a}{2},\frac{a}{2}\right)+C_2^*C_3 G_{\omega_n}^{2,1}\left(\frac{a}{2},\frac{a}{2}\right)-C_3^*C_2 G_{\omega_n}^{2,1}\left(-\frac{a}{2},-\frac{a}{2}\right)-C_1^*C_4
G_{\omega_n}^{2,1}\left(\frac{a}{2},\frac{a}{2}\right)\nonumber\\
&+C_3^*C_2 G_{\omega_n}^{1,2}\left(\frac{a}{2},\frac{a}{2}\right)-
C_2^*C_3 G_{\omega_n}^{1,2}\left(-\frac{a}{2},-\frac{a}{2}\right)+
C_1^*C_4 G_{\omega_n}^{1,2}\left(\frac{a}{2},\frac{a}{2}\right)-
G_{\omega_n}^{2,2}\left(-\frac{a}{2},-\frac{a}{2}\right)\nonumber\\
&+|C_3|^2 G_{\omega_n}^{2,2}\left(\frac{a}{2},\frac{a}{2}\right)-
|C_2|^2 G_{\omega_n}^{2,2}\left(-\frac{a}{2},-\frac{a}{2}\right)+R G_{\omega_n}^{2,2}\left(\frac{a}{2},\frac{a}{2}\right)\bigg].
\label{m15}
\end{align}
Thus, an expression for the current density at the junction is obtained.

\section{Boundary conditions}

The four equations~(\ref{m10}) can be separated into two independent systems of differential equations: the first system of two differential equations for the Green’s functions  $\hat{\mathcal{G}}^{1,1}(t,t')$  and  $\hat{\mathcal{G}}^{2,1}(t,t')$; the second system of two differential equations for the Green’s functions  $\hat{\mathcal{G}}^{2,2}(t,t')$ and  $\hat{\mathcal{G}}^{1,2}(t,t')$. It is significant to note that the first system can be transformed into the second system by the substitutions  $\varphi \rightarrow -\varphi$,  $\hat{\mathcal{G}}^{1,1}(t,t')\rightarrow \hat{\mathcal{G}}^{2,2}(t,t')$ and $\hat{\mathcal{G}}^{2,1}(t,t')\rightarrow \hat{\mathcal{G}}^{1,2}(t,t')$.

Let us consider a system of two differential equations for the Green’s functions  $\hat{\mathcal{G}}^{1,1}(t,t')$ and  $\hat{\mathcal{G}}^{2,1}(t,t')$. In the case $t<-{a}/{2}$, we may write that

\begin{align}
\left\{\begin{array}{ll} \left(\frac{\partial}{\partial t}+\hat{A}(-\varphi) \right)\ri \hat{\mathcal{G}}^{1,1}(t,t')=\delta(t-t'), \\
\left(\frac{\partial}{\partial t}+\hat{A}(\varphi) \right)\ri \hat{\mathcal{G}}^{2,1}(t,t')=0. \end{array}\right.
\label{m16}
\end{align}

In the formula~(\ref{m16}), a notation for a matrix $\hat{A}(\varphi)=\omega_n\sigma_z-\Delta \sigma_y \cos\left({\varphi}/{2}\right)-\Delta \sigma_x \sin({\varphi}/{2})$  is introduced. We can also derive a useful identity
$(\hat{A}(\varphi))^2=\tilde{\omega}_n^2$  containing a notation for a frequency  $\tilde{\omega}_n=\sqrt{\omega_n^2+\Delta^2}$. The system of two differential equations~(\ref{m16}) has a general solution presented by the formula
\begin{align}
\begin{array}{ll}
\hat{\mathcal{G}}^{2,1}(t,t')=\re^{\tilde{\omega}_n\left(t+\frac{a}{2}\right)}
\begin{pmatrix} \alpha_1(t')&\frac{1}{b}\beta_1(t')\\b\alpha_1(t')&\beta_1(t')\\ \end{pmatrix},
\\
\hat{\mathcal{G}}^{1,1}(t,t')=\re^{\tilde{\omega}_n\left(t+\frac{a}{2}\right)}
\begin{pmatrix} \gamma_1(t')&-\frac{1}{b^*}\delta_1(t')\\-b^*\gamma_1(t')&\delta_1(t')\\ \end{pmatrix}
+\frac{1}{2\ri \tilde{\omega}_n}\left[ \tilde{\omega}_n \textrm{sign}(t-t')+\hat{A}(-\varphi) \right]
\re^{-\tilde{\omega}_n|t-t'|}.
\\
\end{array}
\label{m17}
\end{align}

In the formula~(\ref{m17}), a notation for a complex number  $b=\ri \frac{\tilde{\omega}_n+\omega_n}{\Delta}\exp\left[{-\ri \left({\varphi}/{2}\right)}\right]$ is introduced. Considering the case $t>{a}/{2}$, we can derive a system of two differential equations
\begin{align}
\left\{\begin{array}{ll}
\left(\ri \frac{\partial}{\partial t}+\ri \omega_n \sigma_z-\hat{\Delta}_{\varphi}+2\ri \Delta R \sigma_x
\sin\frac{\varphi}{2}\right)\hat{\mathcal{G}}^{1,1}(t,t')-2\ri \Delta C_4^* C_1 \sigma_x \sin\frac{\varphi}{2}\hat{\mathcal{G}}^{2,1}(t,t')=\delta(t-t'),\\
\left(\ri \frac{\partial}{\partial t}+\ri \omega_n \sigma_z-\hat{\Delta}_{-\varphi}-2\ri \Delta R \sigma_x
\sin\frac{\varphi}{2}\right)\hat{\mathcal{G}}^{2,1}(t,t')+2\ri \Delta C_4^* C_1 \sigma_x \sin\frac{\varphi}{2}\hat{\mathcal{G}}^{1,1}(t,t')=0.
\\
\end{array}\right.
\label{m18}
\end{align}
In the formula~(\ref{m18}), a notation for a matrix $\hat{\Delta}_{\varphi}=\ri \Delta \left[\sigma_x \sin \left({\varphi}/{2}\right)+\sigma_y \cos \left({\varphi}/{2}\right) \right]$   is introduced. The system of two differential equations~(\ref{m18}) has a general solution presented by the following formulae:
\begin{align}
\hat{\mathcal{G}}^{1,1}(t,t')&=\re^{-\tilde{\omega}_n\left(t-\frac{a}{2}\right)}
\begin{pmatrix} D\alpha_2(t')+R\gamma_2(t')&R b \delta_2(t')-Db^*\beta_2(t')\\
\frac{R}{b}\gamma_2(t')-\frac{D}{b^*}\alpha_2(t') & D\beta_2(t')+R\delta_2(t')\\ \end{pmatrix} \nonumber\\
&+\frac{1}{2\ri \tilde{\omega}_n}\left[\tilde{\omega}_n \textrm{sign}(t-t')+D\hat{A}(\varphi)+R\hat{A}(-\varphi) \right] \re^{-\tilde{\omega}_n |t-t'|},
\label{m19}
\end{align}
\begin{align}
\hat{\mathcal{G}}^{2,1}(t,t')&=C_1^*C_4\re^{-\tilde{\omega}_n\left(t-\frac{a}{2}\right)}
\begin{pmatrix} \alpha_2(t')-\gamma_2(t')&-b^* \beta_2(t')-b\delta_2(t')\\
-\frac{1}{b^*}\alpha_2(t')-\frac{1}{b}\gamma_2(t') & \beta_2(t')-\delta_2(t')\\ \end{pmatrix} \nonumber\\
&+\frac{C_1^*C_4}{2\ri \tilde{\omega}_n}\left[\hat{A}(\varphi)-\hat{A}(-\varphi) \right] \re^{-\tilde{\omega}_n |t-t'|}.
\label{m20}
\end{align}
In the case $|t|<{a}/{2}$, we have a system of two differential equations
\begin{align}
\left\{\begin{array}{l}
\left(\ri \frac{\partial}{\partial t}+\ri \omega_n \sigma_z-
\left(|C_2|^2+|C_3|^2 \right)\ri \sigma_y\Delta_1 \right)\hat{\mathcal{G}}^{1,1}(t,t')-
(C_2C_3^*+C_2^*C_3)\ri \sigma_y \Delta_1 \hat{\mathcal{G}}^{2,1}(t,t')=\delta(t-t'),\\
\left(\ri \frac{\partial}{\partial t}+\ri \omega_n \sigma_z-
\left(|C_2|^2+|C_3|^2 \right)\ri \sigma_y\Delta_1 \right)\hat{\mathcal{G}}^{2,1}(t,t')-
(C_2C_3^*+C_2^*C_3)\ri \sigma_y \Delta_1 \hat{\mathcal{G}}^{1,1}(t,t')=0.\\
\end{array}\right.
\label{m21}
\end{align}
The system of two differential equations (\ref{m21}) has a general solution presented by the formula
\begin{align}
\hat{\mathcal{G}}^{1,1}(t,t')+(-1)^k \hat{\mathcal{G}}^{2,1}(t,t')&=\re^{-\hat{A}_k(\omega_n)\left(t+\frac{a}{2}\right)}
\left[
\hat{\mathcal{G}}^{1,1}\left(-\frac{a}{2},t'\right)+(-1)^k\hat{\mathcal{G}}^{2,1}\left(-\frac{a}{2},t'\right)
\right]\nonumber\\
&+\frac{1}{2\ri}\left[
\textrm{sign}\left(\frac{a}{2}+t'\right)+\textrm{sign}(t-t')
\right]\re^{-\hat{A}_k(\omega_n)(t-t')}.
\label{m22}
\end{align}
In the formula~(\ref{m22}), the index $k$  can acquire values of  1 and  2. There is a matrix $\hat{A}_k(\omega_n)=\omega_n\sigma_z-|A_k|^2\sigma_y\Delta_1$  containing the notations $A_1=C_2-C_3$  and  $A_2=C_2+C_3$. We can also derive a useful identity  $(\hat{A}_k(\omega_n))^2$ $=(\Omega_k(\omega_n))^2$ containing a notation for the frequency  $\Omega_k(\omega_n)=\sqrt{\omega_n^2+\Delta_1^2|A_k|^4}$. In the matrix equality~(\ref{m22}), the Green’s functions
$\hat{\mathcal{G}}^{1,1}\left(-{a}/{2},t'\right) $ and $\hat{\mathcal{G}}^{2,1}\left(-{a}/{2},t'\right) $  must be calculated using a formula~(\ref{m17}). We can do this, because the Green’s functions are considered to be continuous functions. Considering the elements in the first row and the first column of the matrix equality~(\ref{m22}), we can derive the formula
\begin{align}
&G^{1,1}_{\omega_n}(t,t')+(-1)^k G^{2,1}_{\omega_n}(t,t')=\left\{\cosh\Omega_k\left(t+\frac{a}{2}\right)-\frac{\omega_n}{\Omega_k}
\sinh\Omega_k\left(t+\frac{a}{2}\right) \right\}\nonumber\\
&\times\left\{\gamma_1(t')+(-1)^k\alpha_1(t')+\frac{1}{2\ri \tilde{\omega}_n}\left(\omega_n-\tilde{\omega}_n \textrm{sign}\left(t'+\frac{a}{2}\right) \right)\re^{-\tilde{\omega}_n\left|t'+\frac{a}{2} \right|}
\right\}\nonumber\\
&-\frac{\ri \Delta_1|A_k|^2\sinh\Omega_k\left(t+\frac{a}{2}\right)}{\Omega_k}
\left\{-b^*\gamma_1(t')+(-1)^k b\alpha_1(t')-b^*\frac{\omega_n-\tilde{\omega}_n}{2\ri \tilde{\omega}_n}\re^{-\tilde{\omega}_n\left|t'+\frac{a}{2}\right|} \right\}\nonumber\\
&+\frac{1}{2\ri}\left(\textrm{sign}\left(\frac{a}{2}+t'\right)+\textrm{sign}(t-t') \right)\left\{
\cosh\Omega_k(t-t')-\frac{\omega_n}{\Omega_k}\sinh\Omega_k(t-t')\right\}.
\label{m23}
\end{align}
The unknown constants $\alpha_1(t')$  and $\gamma_1(t')$ should be found with the aim to find the Green’s functions~(\ref{m23}). To do this, a general solution~(\ref{m22}) in the case $|t|<{a}/{2}$ should also be considered on the bound $t={a}/{2}$. As a result, we obtain a boundary condition
\begin{align}
\hat{\mathcal{G}}^{1,1}\left(\frac{a}{2},t'\right)+(-1)^k \hat{\mathcal{G}}^{2,1}\left(\frac{a}{2},t'\right)&=
\re^{-\hat{A}_k(\omega_n)a}
\left[\hat{\mathcal{G}}^{1,1}\left(-\frac{a}{2},t'\right)+(-1)^k \hat{\mathcal{G}}^{2,1}\left(-\frac{a}{2},t'\right)
\right]\nonumber\\
&+\frac{1}{2\ri}
\left[ \textrm{sign}\left(\frac{a}{2}+t'\right)+\textrm{sign}\left(\frac{a}{2}-t'\right)\right]
\re^{-\hat{A}_k(\omega_n)\left(\frac{a}{2}-t' \right)}.
\label{m24}
\end{align}
The Green’s functions  $\hat{\mathcal{G}}^{1,1}\left({a}/{2},t'\right)$ and $\hat{\mathcal{G}}^{2,1}\left({a}/{2},t'\right)$ should be calculated using the formulae~(\ref{m19}) and~(\ref{m20}). The formula~(\ref{m17}) should be used with the aim to calculate the Green’s functions $\hat{\mathcal{G}}^{1,1}\left(-{a}/{2},t'\right)$  and $\hat{\mathcal{G}}^{2,1}\left(-{a}/{2},t'\right)$. Thus, the boundary condition (\ref{m24}) enables us to derive the equalities
\begin{align}
&\left(C_1^*+C_4^*(-1)^k\right)\left[C_4(-1)^k\alpha_2(t')+C_1\gamma_2(t') \right]+
\frac{1}{2\ri \tilde{\omega}_n}\left[\omega_n+\tilde{\omega}_n \textrm{sign}\left(\frac{a}{2}-t'\right) \right]
\re^{-\tilde{\omega}_n\left|t'-\frac{a}{2}\right|}\nonumber\\
&=\left\{\cosh\Omega_ka-\frac{\omega_n}{\Omega_k}\sinh\Omega_ka \right\}
\left\{\gamma_1(t')+(-1)^k\alpha_1(t')+
\frac{1}{2\ri \tilde{\omega}_n}\left[\omega_n-\tilde{\omega}_n \textrm{sign}\left(\frac{a}{2}+t'\right) \right]
\re^{-\tilde{\omega}_n\left|t'+\frac{a}{2}\right|}\right\}\nonumber\\
&-\frac{\ri \Delta_1|A_k|^2\sinh\Omega_ka}{\Omega_k}\left\{
-b^*\gamma_1(t')+(-1)^kb\alpha_1(t')-b^*\frac{\omega_n-\tilde{\omega}_n}{2\ri \tilde{\omega}_n}
\re^{-\tilde{\omega}_n\left|t'+\frac{a}{2}\right|} \right\}\nonumber\\
&+\frac{1}{2\ri}\left[\textrm{sign}\left(\frac{a}{2}+t'\right)+\textrm{sign}\left(\frac{a}{2}-t'\right)   \right]
\left\{ \cosh\Omega_k\left(\frac{a}{2}-t'\right) -\frac{\omega_n}{\Omega_k}\sinh\Omega_k\left(\frac{a}{2}-t'\right)
\right\}
\label{m25}
\end{align}
and
\begin{align}
&\left(C_1+C_4(-1)^{k+1}\right)\left\{
\frac{C_4^*(-1)^k}{b^*}\left(\alpha_2(t')+\frac{\omega_n+\tilde{\omega}_n}{2 \ri \tilde{\omega}_n}
\re^{-\tilde{\omega}_n\left|t'-\frac{a}{2}\right|}\right)+
\frac{C_1^*}{b}\left(\gamma_2(t')+\frac{\omega_n+\tilde{\omega}_n}{2 \ri \tilde{\omega}_n}
\re^{-\tilde{\omega}_n\left|t'-\frac{a}{2}\right|}\right)\right\}\nonumber\\
&=\frac{\ri \Delta_1 |A_k|^2\sinh\Omega_ka}{\Omega_k}
\left\{ \gamma_1(t')+(-1)^k\alpha_1(t')+\frac{1}{2 \ri \tilde{\omega}_n}\left[\omega_n-\tilde{\omega}_n \textrm{sign}\left(\frac{a}{2}+t'\right) \right]
\re^{-\tilde{\omega}_n\left|t'+\frac{a}{2}\right|}\right\}\nonumber\\
&+\left\{ \cosh\Omega_ka+\frac{\omega_n}{\Omega_k}  \sinh\Omega_ka \right\}
\left\{ -b^*\gamma_1(t')+(-1)^kb\alpha_1(t')-
b^*\frac{\omega_n-\tilde{\omega}_n}{2\ri\tilde{\omega}_n}\re^{-\tilde{\omega}_n\left|t'+\frac{a}{2}\right|}
\right\}\nonumber\\
&+\frac{1}{2\ri}\left[\textrm{sign}\left(\frac{a}{2}+t'\right)+\textrm{sign}\left(\frac{a}{2}-t'\right)\right]
\frac{\ri \Delta_1 |A_k|^2\sinh\Omega_k\left(\frac{a}{2}-t'\right)}{\Omega_k}.
\label{m26}
\end{align}
The equalities~(\ref{m25}) and~(\ref{m26}) form a closed system of four linear equations for the following unknowns:  $\alpha_1(t')$,  $\alpha_2(t')$, $\gamma_1(t')$  and  $\gamma_2(t')$. To find the Green’s functions  $G^{1,1}_{\omega_n}(t,t')$ and  $G^{2,1}_{\omega_n}(t,t')$, the calculated constants  $\alpha_1(t')$ and $\gamma_1(t')$ should be substituted into a formula~(\ref{m23}). The other Green’s functions can be found by the substitutions $\varphi\rightarrow -\varphi$, $G^{1,1}_{\omega_n}(t,t')\rightarrow G^{2,2}_{\omega_n}(t,t')$  and  $G^{2,1}_{\omega_n}(t,t')\rightarrow G^{1,2}_{\omega_n}(t,t')$.

\section{Current-phase relation}

The four Green’s functions should be calculated in the cases $t=t'=-{a}/{2}$  and $t=t'={a}/{2}$. The obtained results should be substituted into the formula~(\ref{m15}). As a result, we obtain a dependence of the current density on the phase difference in the layered superconducting structures of a SIS’IS type
\begin{align}
j\left(\frac{d}{2}\right)=\piup e v_0 N(0)T\sum_{\omega_n}\int_0^1
\frac{\left[D\sin \varphi+\frac{1+D}{8}E_{\omega_n}(x,a,\varphi) \right]\Delta^2x\rd x}
{\omega_n^2+\Delta^2\left[1-D\sin^2\frac{\varphi}{2}\right]+F_{\omega_n}(x,a,\varphi)}.
\label{m27}
\end{align}
In the formula~(\ref{m27}), there are notations
\begin{align}
&E_{\omega_n}(x,a,\varphi)=2(\cosh\Omega_1a\cosh\Omega_2a-1)\sin \varphi\nonumber\\
&+\frac{4\Delta_1}{\Delta}|A_1|^2\tilde{\omega}_n\frac{\sinh\Omega_1a\cosh\Omega_2a}{\Omega_1}\sin\frac{\varphi}{2}+
\frac{4\Delta_1}{\Delta}|A_2|^2\tilde{\omega}_n\frac{\cosh\Omega_1a\sinh\Omega_2a}{\Omega_2}\sin\frac{\varphi}{2}
\nonumber\\
&+\left[ \frac{4\Delta_1}{\Delta}\left(|A_1|^2+|A_2|^2 \right)\omega_n^2\sin\frac{\varphi}{2}-
2\left(\omega_n^2-\Delta_1^2|A_1|^2|A_2|^2  \right)\sin\varphi\right]
\frac{\sinh\Omega_1a\sinh\Omega_2a}{\Omega_1\Omega_2},
\label{m28}
\end{align}
\begin{align}
&F_{\omega_n}(x,a,\varphi)=\left\{\omega_n^2+\frac{\Delta^2}{2}\left[1+\cos^2\frac{\varphi}{2}\right]\right\}
\left(\cosh\Omega_1a\cosh\Omega_2a-1\right)\nonumber\\
&+\frac{\tilde{\omega}_n}{\Omega_1}\left(\omega_n^2+\Delta_1\Delta|A_1|^2\cos\frac{\varphi}{2}\right)
\sinh\Omega_1a\cosh\Omega_2a+
\frac{\tilde{\omega}_n}{\Omega_2}\left(\omega_n^2+\Delta_1\Delta|A_2|^2\cos\frac{\varphi}{2}\right)
\cosh\Omega_1a\sinh\Omega_2a\nonumber\\
&+\frac{1}{\Omega_1\Omega_2}\Big\{
\left(\omega_n^2+\Delta_1\Delta|A_1|^2\cos\frac{\varphi}{2}\right)
\left(\omega_n^2+\Delta_1\Delta|A_2|^2\cos\frac{\varphi}{2}\right)\nonumber\\
&+\frac{\Delta^2}{2}\left(\omega_n^2+\Delta_1^2|A_1|^2|A_2|^2\right)\sin^2\frac{\varphi}{2}\Big\}
\sinh\Omega_1a\sinh\Omega_2a.
\label{m29}
\end{align}

It is significant to note that the current-phase relation~(\ref{m27}) is the main result of our investigation. It was found that the formula~(\ref{m27}) can be used with the aim to consider the particular cases.

\begin{figure}[!t]
\centerline{\includegraphics[width=0.75\textwidth]{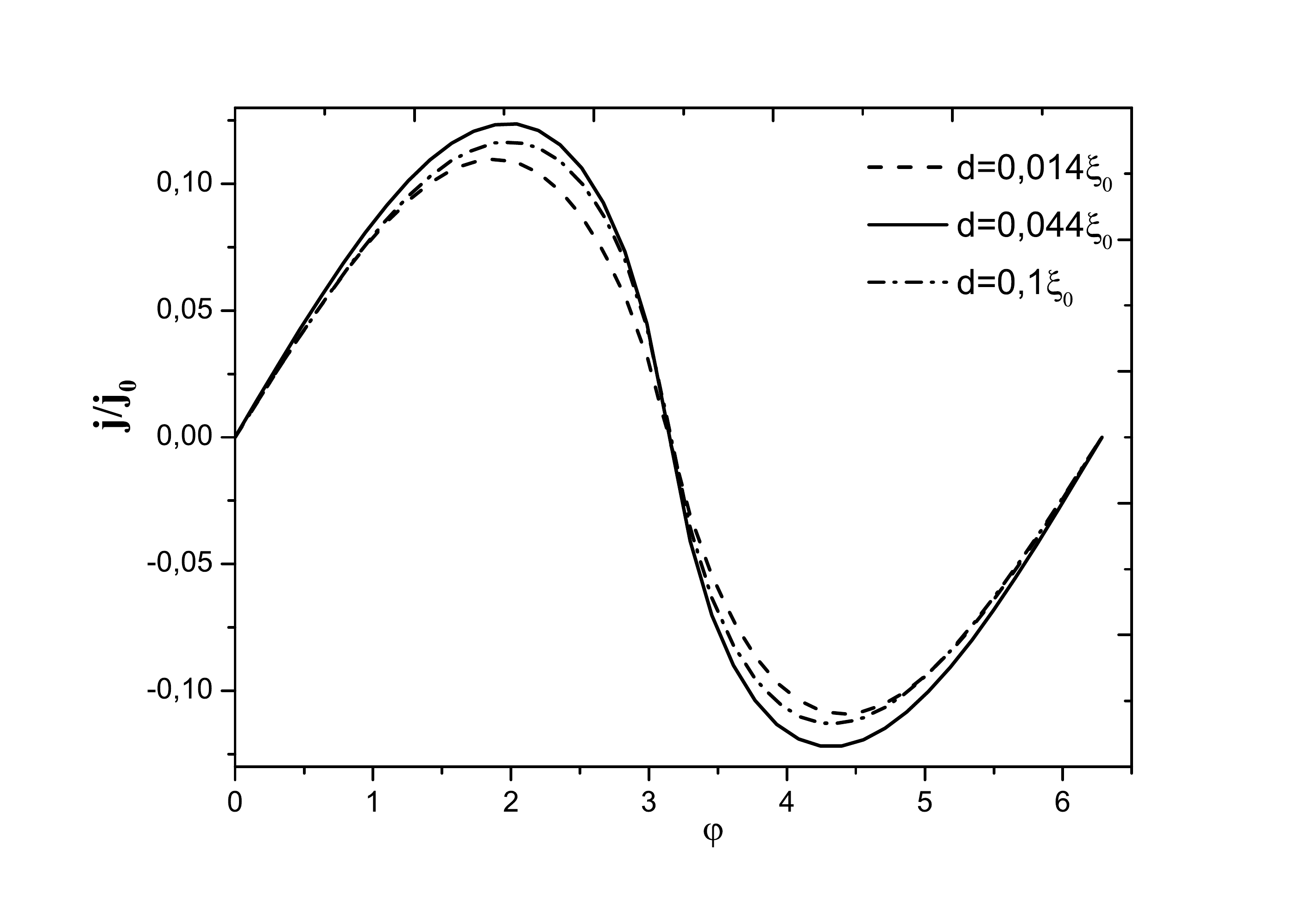}}
\caption{The current-phase relation for the temperature $T=0.5T'_{c}$ and the dielectric layer transparency $D=0.1$ with various values of an intermediate layer thickness $d$. $T'_{c}$ is the critical temperature of an intermediate layer S'.} 
\label{fig-smp1}
\end{figure}

\begin{figure}[!t]
\centerline{\includegraphics[width=0.75\textwidth]{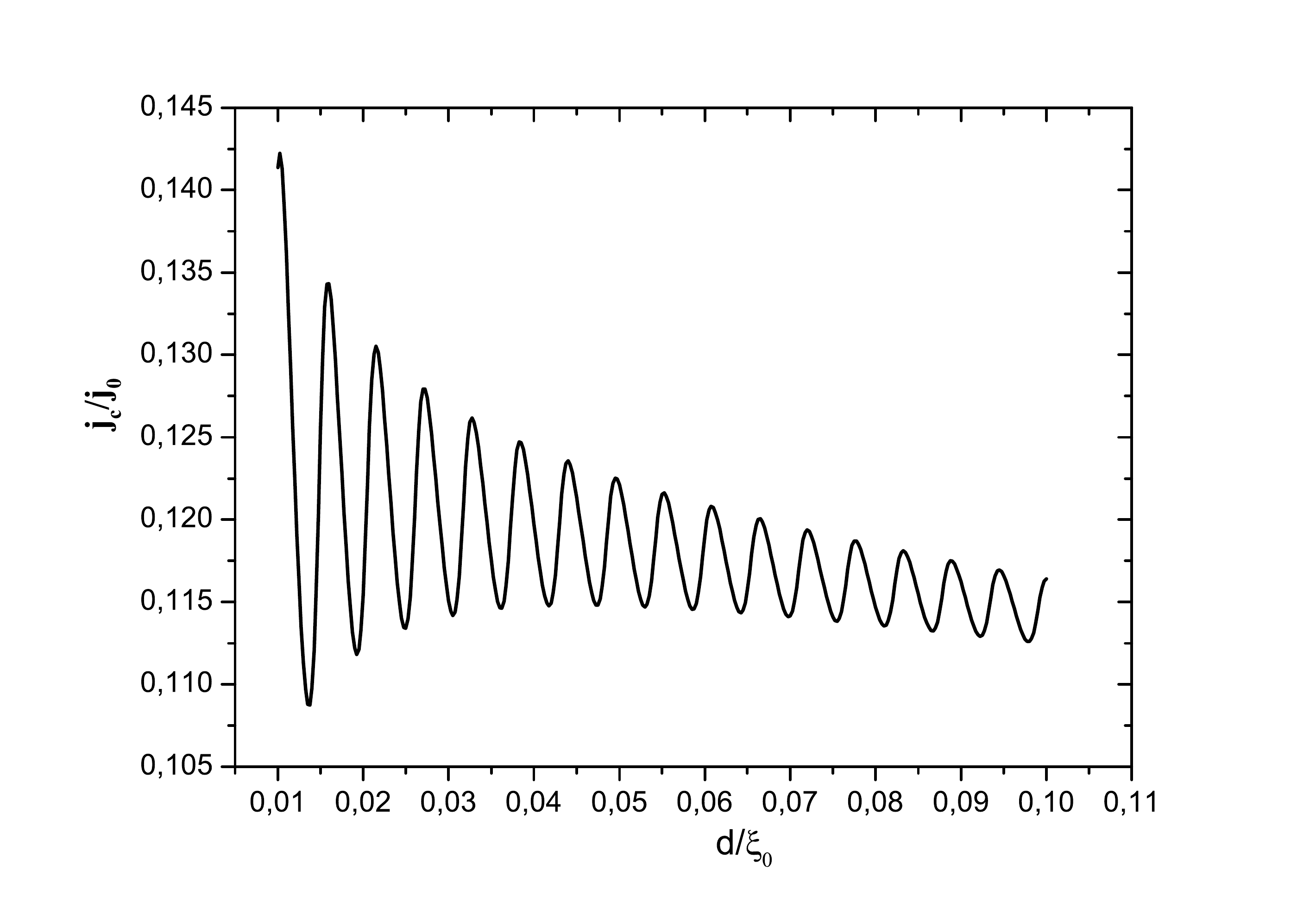}}
\caption{The dependence of the dimensionless critical current ${j_c}/{{{j}_{0}}}$ on the thickness $d$ of the intermediate layer~S'.} 
\label{fig-smp2}
\end{figure}

\begin{figure}[!t]
\centerline{\includegraphics[width=0.75\textwidth]{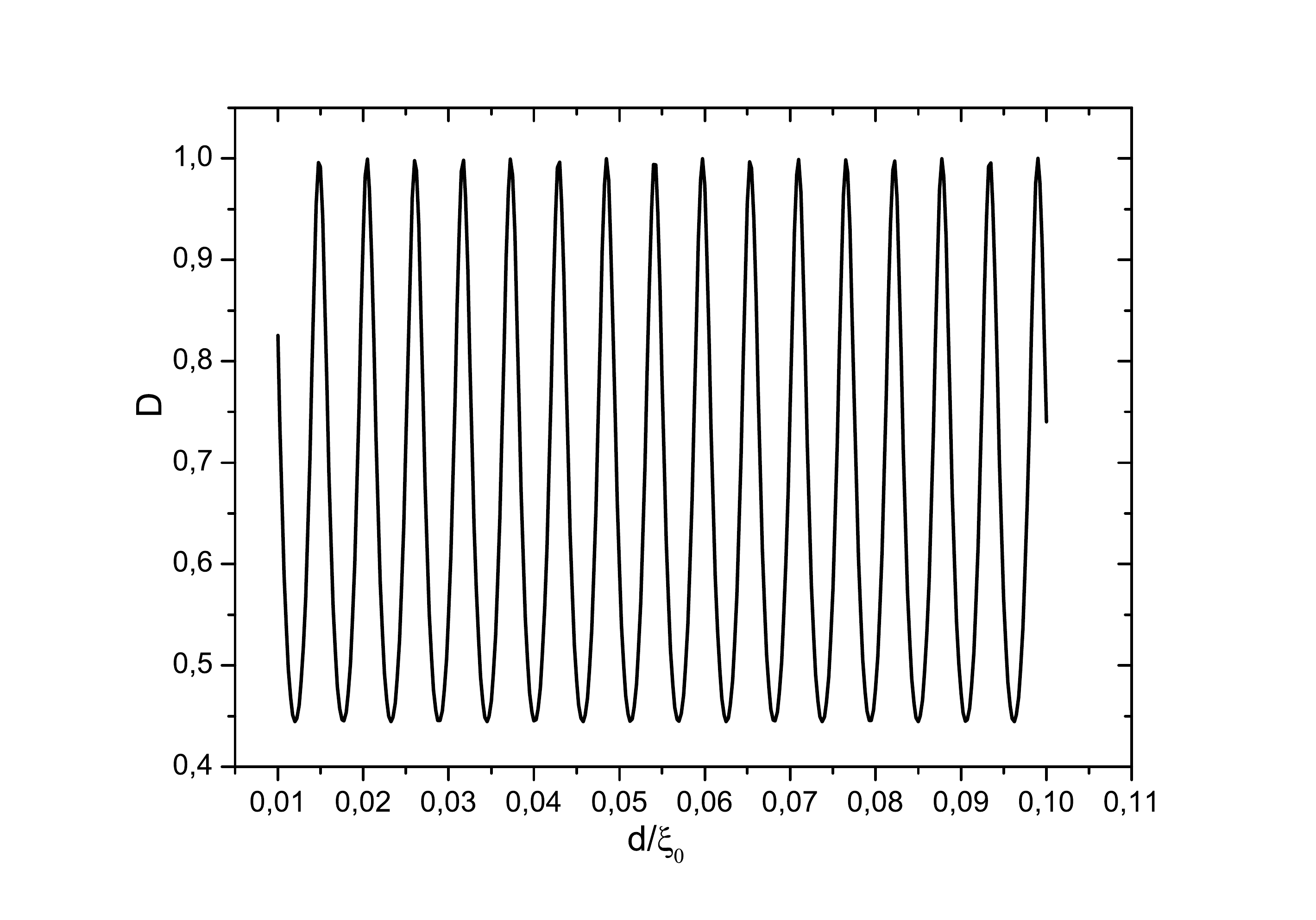}}
\caption{The dependence of the dielectric layer transparency $D$ on the thickness $d$ of the intermediate layer~S'.} 
\label{fig-smp3}
\end{figure}

In figure~\ref{fig-smp1}, the dependence of the dimensionless current density ${j}/{{{j}_{0}}}$ on a phase difference $\varphi $ in the layered superconducting structures of a SIS’IS type is presented using the current-phase relation~(\ref{m27}) obtained in our investigation. All required calculations are performed considering the relationship ${{T}_{c}}~=~7.4{{T'_c}}$ between the critical temperatures ${{T}_{c}}$ and ${{T}'_{c}}$. It is significant to note that the above presented relationship is a relationship between the critical temperatures of niobium (Nb) and aluminium (Al) in the junctions of a Nb|Al|Nb type. The current density $j$ is also dependent on the thickness $d$ of an intermediate layer S'. The particular cases $d=0.014{{\xi }_{0}}$, $d=0.044{{\xi }_{0}}$ and $d=0.1{{\xi }_{0}}$ are analyzed with the aim to obtain the figure~\ref{fig-smp1}. A common feature of the three intermediate layer thicknesses is the dimensionless current density ${j}/{{{j}_{0}}}$ with the maximum shifted into the area $\varphi >{\piup }/{2}$. It is also important to note that the absolute value of the critical current in the case of the thickness $d=0.044{{\xi }_{0}}$ is higher than the corresponding absolute value of the critical current in case of the thickness $d=0.014{{\xi }_{0}}$. This interesting fact can be explained by the dependence of the dielectric layer transparency $D$ on the thickness $d$ of the intermediate layer~S' because the above-mentioned dependence is a periodic function. For certain values of the intermediate layer thickness $d$, the dielectric layer transparency $D$ can be equal to  1 (see figure~\ref{fig-smp2}). These are the so-called resonant tunneling modes in the layered superconducting structures of a SIS'IS type. The most striking manifestation of the resonant tunneling modes can be observed in figure~\ref{fig-smp3} containing the dependence of the critical current at the junction on the thickness $d$ of the intermediate layer S'. The current density $j$ is presented via the dielectric layer transparency $D$. However, the current density $j$ is also presented via the functions ${{E}_{{{\omega }_{n}}}}\left( x,a,\varphi  \right)$ and ${{F}_{{{\omega }_{n}}}}\left( x,a,\varphi  \right)$. As a result, the damped oscillation of the critical current can be observed. When the thickness $d$ of the intermediate layer S' is increased, the average value of the critical current decreases.

Let us consider a particular case  $\Delta_1=0$. As a result, the superconducting junction of a SINIS type is obtained. Formula~(\ref{m27}) can be transformed into the formula
\begin{align}
j\left(\frac{d}{2}\right)=2\piup e v_0 N(0)T\sum_{\omega_n}\int_0^1
\frac{\Delta^2D\sin\varphi}
{(\tilde{\omega}_n^2+\omega_n^2)\cosh2\omega_na+2\tilde{\omega}_n\omega_n\sinh2\omega_na+
\Delta^2(R+D\cos\varphi)}x\rd x.
\label{m30}
\end{align}

In the asymptotical case $R\ll 1$, formula~(\ref{m30}) can also be considered. This enables us to neglect the reflection coefficient $R$. We obtain a superconducting junction of a SNS type. The current-phase relation is defined by the formula
\begin{align}
j\left(\frac{d}{2}\right)=2\piup e v_0 N(0)T\sum_{\omega_n}\int_0^1
\frac{\Delta^2\sin\varphi}
{(\tilde{\omega}_n^2+\omega_n^2)\cosh2\omega_na+2\tilde{\omega}_n\omega_n\sinh2\omega_na+
\Delta^2\cos\varphi}x\rd x.
\nonumber
\end{align}
In the asymptotical case $d\gg\xi_0$, formula~(\ref{m27}) can be transformed into the formula
\begin{align}
j\left(\frac{d}{2}\right)=\frac{\piup \Delta^2}{8} e v_0 N(0)T\sum_{\omega_n}\int_0^1x(1+D)
\frac{\tilde{E}_{\omega_n}(x,a,\varphi)}{\tilde{F}_{\omega_n}(x,a,\varphi)}\rd x
\nonumber
\end{align}
containing the notations
\begin{align}
\tilde{E}_{\omega_n}(x,a,\varphi)&=2\sin \varphi+\frac{4\Delta_1}{\Delta}
\left(\frac{|A_1|^2}{\Omega_1}+\frac{|A_2|^2}{\Omega_2}\right)\tilde{\omega}_n\sin\frac{\varphi}{2}
\nonumber\\
&+\frac{1}{\Omega_1\Omega_2}\left[
\frac{4\Delta_1}{\Delta}\left(|A_1|^2+|A_2|^2\right)\omega_n^2\sin\frac{\varphi}{2}-
2\left(\omega_n^2-\Delta_1^2|A_1|^2|A_2|^2\right)\sin\varphi \right],\nonumber
\end{align}
\begin{align}
\tilde{F}_{\omega_n}(x,a,\varphi)=\omega_n^2+\frac{\Delta^2}{2}
\left[1+\cos^2\frac{\varphi}{2}\right]+
\frac{\tilde{\omega}_n}{\Omega_1}\left(\omega_n^2+\Delta_1\Delta|A_1|^2 \cos\frac{\varphi}{2}\right)+
\frac{\tilde{\omega}_n}{\Omega_2}\left(\omega_n^2+\Delta_1\Delta|A_2|^2 \cos\frac{\varphi}{2}\right)
\nonumber\\
+\frac{1}{\Omega_1\Omega_2}\left\{
\left(\omega_n^2+\Delta_1\Delta|A_1|^2 \cos\frac{\varphi}{2}\right)
\left(\omega_n^2+\Delta_1\Delta|A_2|^2 \cos\frac{\varphi}{2}\right)+
\frac{\Delta^2}{2}\left(\omega_n^2+\Delta_1^2|A_1|^2|A_2|^2\right)\sin^2\frac{\varphi}{2}\right\}.\nonumber
\end{align}

In the asymptotical case $d\ll \xi_0$, the functions~(\ref{m28}) and~(\ref{m29}) are considered to be infinitesimal quantities. The approximations  $E_{\omega_n}(x,a,\varphi)\cong 0$ and $F_{\omega_n}(x,a,\varphi)\cong 0$  can be used with the aim to transform formula~(\ref{m27}). A well-known identity
\[T\sum_{\omega_n}\frac{1}{\omega_n^2+\alpha^2}=\frac{1}{2\alpha}\tanh\frac{\alpha}{2T}\]
enables us to derive the current-phase relation~\cite{s18}
\[
j\left(\frac{d}{2}\right)=\frac{\piup}{2} e v_0 N(0)\int_0^1\left(
\frac{\Delta x D(x) \rd x}{\sqrt{1-D(x)\sin^2\frac{\varphi}{2}}}
\tanh\frac{\Delta\sqrt{1-D(x)\sin^2\frac{\varphi}{2}}}{2T}
\right)\sin\varphi.
\]

\section{Conclusions}

In this paper, a theoretical study is devoted to considering a three-dimensional superconducting junction of a SIS’IS type. It is significant to note that an order parameter is considered to be a piecewise constant function. It is understood that a model with a piecewise constant order parameter cannot be applicable at the temperatures close to the critical temperature $T_c$. The spatial behavior of an order parameter must always be considered at the temperatures close to the critical temperature $T_c$. As a result, a self-consistency condition must be applied. Using the Matsubara Green’s functions, a mathematical description of a three-dimensional SIS’IS junction is realized by the second order differential equations that are also known as the Gorkov’s equations. It was found that the model with a piecewise constant order parameter enables one to transform the Gorkov’s equations into the quasiclassical equations for the Green’s functions in a $t$-representation. It is significant to note that the quasiclassical equations are the first order differential equations. This is the main advantage of quasiclassical equations. The proposed calculation scheme enables one to obtain a new analytical expression for the dependence of the current density on the phase difference. It was also found that a new current-phase relation is much different from a sinusoidal current-phase relation. It is also important to note that a new current-phase relation is true for arbitrary values of the dielectric layer transparency and the intermediate layer thickness.

\newpage
\ukrainianpart

\title{Струм-фазова залежність у шаруватих надпровідних структурах типу SIS’IS}
\author{А. М. Шутовський, В. Є. Сахнюк}
\address{Волинський національний університет імені Лесі Українки, просп. Волі, 13, 43000 Луцьк, Україна}

\makeukrtitle

\begin{abstract}
\tolerance=3000%
Досліджено залежність густини струму від різниці фаз, розглядаючи шаруваті надпровідні структури типу SIS’IS. Побудовано квазікласичні рівняння для функцій Гріна в $t$-представленні з метою спрощення обрахунків. Параметр впорядкування подано у вигляді кусково-сталої функції. Розглянуто загальний випадок, не накладаючи ніяких обмежень на прозорість діелектричного прошарку та товщину проміжного надпровідника. Виявилося, що новий аналітичний вираз для струм-фазової залежності містить у собі низку раніше відомих результатів для частинних випадків.

\keywords функція Гріна, параметр впорядкування, густина струму, різниця фаз

\end{abstract}


\begin{thebibliography}{10}

\bibitem{r09} Radovi\'c Z., Paltoglou V., Lazarides N., Flytzanis N., Eur. Phys. J. B, 2009, \textbf{69}, 229--236,\\
 \doi{10.1140/epjb/e2009-00133-4}.
\bibitem{n99} Nevirkovets I. P., Shafranjuk S. E., Phys. Rev. B, 1999,  \textbf{59}, 1311--1317,
 \doi{10.1103/PhysRevB.59.1311}.
\bibitem{b00} Brinkman A., Golubov A. A., Phys. Rev. B, 2000, \textbf{61}, 11297--11300,
 \doi{10.1103/PhysRevB.61.11297}.
\bibitem{f65} Feynman R. P., Leighton R. B., Sands M., Feynman Lectures on Physics, Vol.~3,
Addison-Wesley, \\Boston, 1965.
\bibitem{b82} Barone A., Paterno G., Physics and Applications of the Josephson Effect, Wiley, New York, 1982.
\bibitem{m02} Mei T., Int. J. Mod. Phys. B, 2002, \textbf{16}, No.~24, 3697--3705,
 \doi{10.1142/S0217979202013080}.
\bibitem{c96} Carapella G., Costabile G., de Luca R., Pace S., Polcari A., Soriano C., Physica C, 1996, \\ \textbf{259}, 349--355,
 \doi{10.1016/0921-4534(96)00115-3}.
\bibitem{o76} Ohta H., A Self-Consistent Model of the Josephson Gunction, IC-SQUID~76, Walter de Gruyter, \\ Berlin, 1976, 35--49.
\bibitem{s06} Shafranjuk S. E., Phys. Rev. B, 2006, \textbf{74}, 024521 (9 pages),
 \doi{10.1103/PhysRevB.74.024521}.
\bibitem{p00} Pepe G. P., Ammendola G., Peluso G., Barone A., Appl. Phys. Lett., 2000, \textbf{77}, 447--449,\\
 \doi{10.1063/1.127005}.
\bibitem{l13} De Luca R., Eur. Phys. J. B, 2013, \textbf{86}, No.~6, 294 (8 pages),
 \doi{10.1140/epjb/e2013-40095-2}.
\bibitem{n94} Nevirkovets I. P., Evetts J. E., Blamire M. G., Phys. Lett. A, 1994, \textbf{187}, No.~1, 119--126,\\
 \doi{10.1016/0375-9601(94)90876-1}.
\bibitem{k99} Kupriyanov M.~Yu., Brinkman A., Golubov A. A., Siegel M., Rogalla H., Physica C, 1999, \textbf{326--327},\\ 16–45,
 \doi{10.1016/S0921-4534(99)00408-6}.
\bibitem{n999} Nevirkovets I. P., Ketterson J. B., Lomatch S., Appl. Phys. Lett., 1999, \textbf{74},  No.~11, 1624--1626,\\
 \doi{10.1063/1.123637}.
\bibitem{s98} Schulze H., Behr R., M\"uller F., Niemeyer J., Appl. Phys. Lett., 1998, \textbf{73}, No.~7, 996--998,\\
 \doi{10.1063/1.122064}.
\bibitem{n97} Nevirkovets I. P., Evettes J. E., Blamire M. G., Barber Z. H., Goldobin E., Phys. Lett. A, 1997, \textbf{232}, \\299--304,
 \doi{10.1016/S0375-9601(97)00388-5}.
\bibitem{a66} Andreev A. F., Sov. Phys. JETP, 1966, \textbf{22}, No.~2, 455--458.
\bibitem{s82} Svidzinskyi A. V., Spatially Inhomogeneous Problems in the Theory of Superconductivity, Nauka, \\Moscow, 1982, (in Russian).
\bibitem{p11}  Pastukh O. Y., Sakhnyuk V. E.,  Svidzinsky A. V., Phys. Lett. A, 2018, \textbf{382}, 2149--2155,\\
\doi{10.1016/j.physleta.2018.05.035}
\bibitem{p18} Pastukha O. Yu., Shutovskii A. M., Sakhnyuk V. E. Low Temp. Phys., 2017, \textbf{43}, No.~6, 664--669,\\
\doi{10.1063/1.4985972}
\bibitem{s18} Shygorin P., Svidzynskyi A., Materian I., Ukr. J. Phys., 2018, \textbf{62}, No.~6, 518--525,\\
 \doi{10.15407/ujpe62.06.0518}.

\end{thebibliography}
\end{document}